\documentclass[preprint,showpacs,showkeys] {revtex4}

\usepackage{graphicx}
\usepackage{bm}
\usepackage{amsfonts,amsmath,amssymb}
\usepackage[squaren]{SIunits}

\begin{document}

\title{Intergrowth and thermoelectric properties in the Bi-Ca-Co-O system}
\author{X. G. Luo, Y. C. Jing, H. Chen, and X. H. Chen\footnote{Corresponding Author. Tex/Fax:+86 551 3601654.\\ \emph{E-mail address} : chenxh@ustc.edu.cn (X.H. Chen)}}
\affiliation{Hefei National Laboratory for Physical Science at
Microscale and Department of Physics, University of Science and
Technology of China, Hefei, Anhui 230026, People's Republic of
China}

\begin{abstract}
Single crystals of the Bi-Ca-Co-O system have been grown using the
flux method with cooling from 900$\celsius$ and 950$\celsius$,
respectively. The single crystals are characterized by transmission
electron microscopy and X-ray diffraction. The misfit cobaltite
[Ca$_2$Bi$_{1.4}$Co$_{0.6}$O$_4$]$^{RS}$[CoO$_2$]$_{1.69}$ single
crystals with quadruple ($n$=4) rocksalt (RS) layer are achieved
with cooling from 900$\celsius$. Such crystal exhibits
room-temperature thermoelectric power (TEP) of 180$\mu$V/K, much
larger than that in Sr-based misfit cobaltites with quadruple RS
layer. However, intergrowth of single crystals of quadruple ($n$=4)
and triple ($n$=3) RS-type layer-based misfit cobaltites is observed
with cooling from 950$\celsius$. Both of TEP and resistivity were
obviously enhanced by the intergrowth compared to
[Ca$_2$Bi$_{1.4}$Co$_{0.6}$O$_4$]$^{RS}$[CoO$_2$]$_{1.69}$ single
crystal, while the power factor at room temperature remains
unchanged.
\end{abstract}


\maketitle

\section{INTRONDUCTION}

The first cobaltite Na$_x$CoO$_2$ exhibiting large TEP ($S \sim$ 100
$\mu$V/K at room temperature) as well as low electrical resistivity
was discovered by Terasaki \textit{et al.} in 1997 \cite{Terasaki}.
Since then, the misfit cobaltites have also been thought to be the
potential candidates for thermoelectric applications. The crystal
structure of the misfit cobaltites consists of alternatively
stacking three ($n$=3, such as
[Ca$_2$CoO$_3$]$^{RS}$[CoO$_2$]$_{1.62}$) or four ($n$=4, such as
namely [$M_2$Bi$_2$O$_4$]$^{RS}$[CoO$_2$]$_{\alpha}$, $M$=Ca, Sr,
Ba) RS-type layers and one CoO$_2$ hexagonal CdI$_2$-type layer with
edge-shared CoO$_6$ octahedra \cite{Masset,Leligny}, which is
similar to that found in Na$_x$CoO$_2$. For simplicity, we take them
as $n$=3 phase and $n$=4 phase, respectively. The two sublattices of
rocksalt block and hexagonal $CdI_2$-type $CoO_2$ layer possess the
common $a$- and $c$-axis lattice parameters and $\beta$ angles but
different $b$-axis length causing a misfit along $b$-direction.
Among all the misfit cobaltites, TEP varies from $S$ = 90 $\mu$V/K
for [Sr$_2$Co$_{1-x}$Tl$_x$O$_3$]$^{RS}$[CoO$_2$]$_{1.8}$ up to $S$
= 165 $\mu$V/K for
[Ca$_2$Co$_{0.6}$Pb$_{0.4}$O$_3$]$^{RS}$[CoO$_2$]$_{1.61}$
\cite{Maignan1,Maignan2,Maignan4}. Bi$^{3+}$ seems to be the most
beneficial to be included in these cobaltites in terms of the
thermoelectric figure of merit ($ZT=S^2T/\rho\kappa$, $S$,$\rho$ and
$\kappa$ are TEP, electrical resistivity and thermal conductivity,
respectively) and is now used in the development for oxide-based
thermogenerators \cite{Funahashi}. In the
[Ca$_2$CoO$_3$]$^{RS}$[CoO$_2$]$_{1.62}$ ($n$=3) cobaltite, its
room-temperature TEP value increases from 125 to 140 $\mu$V/K by low
bismuth amounts substitution for Ca \cite{Mikami}. The
polycrystalline
[Bi$_{1.7}$Co$_{0.3}$Ca$_2$O$_4$]$^{RS}$[CoO$_2$]$_{1.67}$ ($n$=4)
\cite{Maignan3} with large amounts of bismuth shows about 140
$\mu$V/K at room temperature. The RS block in
[Bi$_{1.7}$Co$_{0.3}$Ca$_2$O$_4$]$^{RS}$[CoO$_2$]$_{1.67}$ is
constructed with two deficient [BiO] layers sandwiched by two [CaO]
layers. In contrast, the RS block in
[Ca$_2$CoO$_3$]$^{RS}$[CoO2]$_{1.62}$ is built up from two [CaO]
layers sandwiching one [CoO] layer \cite{Leligny,Maignan3}. Though
the number of the RS layers is different, they show almost the same
$b_{RS}$/$b_{H}$ ratio ($b_{RS}$ and $b_H$ are the lattice
parameters along b-axis for RS and hexagonal layer, respectively)
\cite{Masset}. The common ground between Bi-doped
[Ca$_2$CoO$_3$]$^{RS}$[CoO2]$_{1.62}$ and
[Bi$_{1.7}$Co$_{0.3}$Ca$_2$O$_4$]$^{RS}$[CoO$_2$]$_{1.67}$ evokes
the sufficient interest of the role of $n$=3/$n$=4 RS layer to the
thermoelectric properties. In this article, single crystals of
Bi-Ca-Co-O were grown by the flux method cooling from different
melting temperatures. It is found that single crystals with pure
$n$=4 phase were obtained with cooling from 900$\celsius$, while
intergrowth of single crystals of quadruple ($n$=4) and triple
($n$=3) RS-type layer-based misfit cobaltites were achieved with
cooling from 950$\celsius$. Structural analysis and thermoelectric
properties were systematically studied on these single crystals.

\section{EXPERIMENTAL DETAIL}
The Bi-Ca-Co-O single crystals were grown by the solution method
using K$_2$CO$_3$-KCl fluxes. In the first step, polycrystalline
Bi$_2$Ca$_2$Co$_2$O$_y$ was prepared by a solid-state reaction
method. Starting materials Bi$_2$O$_3$, CaCO$_3$ and Co$_3$O$_4$
were mixed in a proportion of Bi:Ca:Co = 2:2:2, with a total weigh
to be 2.7 grams. The powders were heated at 800$\celsius$ for 10
hours. Then the prepared Bi$_2$Ca$_2$Co$_2$O$_y$ and the mixture of
KCl and K$_2$CO$_3$ by a molar proportion of 1:4 (20.5 grams) were
mixed and loaded in an aluminum crucible having 30 ml volume. The
solute concentration was about 1.5 mol$\%$. A lidded crucible was
used to prevent the solution from evaporating and to grow crystals
under stable conditions. The powder-flux mixture was melted at
900$\celsius$ or 950$\celsius$ for 20 hours, respectively, and then
slowly cooled down to 600$\celsius$ at a rate of 5-6$\celsius$/hr.
The single crystals were separated from the melt by washing with
distilled water. The single crystals obtained with cooling from
900$\celsius$ and 950$\celsius$ were denoted as crystal 1 and
crystal 2 in the following text, respectively. The crystals were
large thin platelets and black in color, as shown in Fig.1. As shown
in Fig.1, typical dimensions of the crystal 1 and crystal 2 are
around 3$\times$3$\times$0.05 mm$^3$ and 5$\times$5$\times$0.05
mm$^3$, respectively.

The structural characterization of the single crystals was performed
by X-ray diffraction (XRD) and transmission electron diffraction
(TEM). The analysis of the actual composition was made by the
inductively coupled plasma (ICP) atomic emission spectroscopy (AES)
(ICP-AES) technique. Resistivity was measured using a standard four
probe method through an alternative current (AC) resistance bridge
(LR700, Linear Research Inc.). The TEP was measured using the
steady-state technique.

\section{RESULTS and DISCUSSION}
\subsection{Structural Characterization}
The XRD patterns recorded for crystal 1 and crystal 2 are shown in
Fig. 2. The presence of only 00$l$ reflections indicates that the
crystals are grown along c-axis. The XRD pattern for crystal 1 grown
from 900$\celsius$ indicates good single $n$=4 phase. The $c$-axis
lattice parameter was estimated to be 14.651 ${\rm \AA}$ based on
the d-spacing values of 00$l$ reflections. The ICP-AES gave that the
actual composition for crystal 1 is Bi : Ca : Co = 1.40 : 2.00 :
2.29. The XRD pattern for crystal 2 shows two sets of 00$l$
reflections. Based on d-spacing values for the two sets of the
reflections, the c-axis lattice parameters for the two sets of 00$l$
reflections are 14.659 $\rm \AA$ and 10.800 ${\rm \AA}$,
respectively. It suggests that there exist two phases in the crystal
2. The 00$l$ reflections arises from the structure of the $n$=4
phase ($c$ = 14.659 ${\rm \AA}$) as the crystal 1, while the set of
00$l$ reflections with c= 10.800 ${\rm \AA}$ is the same as that
observed in [Ca$_2$CoO$_3$]$^{RS}$[CoO$_2$]$_{1.62}$ ($n$=3) phase.
Therefore, it indicates that an intergrowth of the $n$=4 phase and
[Ca$_2$CoO$_3$]$^{RS}$[CoO$_2$]$_{1.62}$ ($n$=3) phase occurs in
crystal 2. Similar intergrowth has been reported in polycrystalline
Sr-Bi-Co-O system, where the major phase is $n$=3 layer-based
cobaltite \cite{Maignan4}. The XRD pattern of crystal 2 shows that
the dominated phase is $n$=4. The ICP-AES gave the actual
composition for crystal 2 to be Bi : Ca : Co = 1.40 : 2.37 : 2.77.

The electron diffraction (ED) patterns are shown in Fig. 3 for the
crystal 1 and crystal 2. In Fig.3a, one can clearly see RS
diffraction spots from the [Bi$_2$Ca$_2$O$_4$] layer and hexagonal
diffraction spots from the [CoO$_2$] layer for crystal 1. The $a$-
and $b$-axis length of the hexagonal [CoO$_2$] layer ($a_H$, $b_H$)
is estimated to be 2.87(9) and 2.83(2) $\AA$, respectively. The $a$-
and $b$-axis length of the RS [Bi$_2$Ca$_2$O$_4$] layer ($a_{RS}$,
$b_{RS}$) is estimated to be 4.98(9) ($\approx \sqrt{3}a_H$)  and
4.784 $\AA$, respectively. From the above structural analysis,
crystal 1 shows the misfit structure along $b$-axis ($b_{RS}/b_{H}$
= 1.69); while along $a$-axis, length of the rock-layer matches with
that of the hexagonal layer ($a_{RS}\approx \sqrt{3}a_{H}$), being
consistent with previous report in polycrystalline sample
\cite{Tanaka}. Therefore, the structural formula of the crystal 1
can be written as
[Ca$_2$Bi$_{1.4}$Co$_{0.6}$O$_4$]$^{RS}$[CoO$_2$]$_{1.69}$.

Fig.3b shows similar main ED pattern for crystal 2 to that observed
in crystal 1 shown in Fig.3a. It gives the same in-plane lattice
parameters of RS and hexagonal layers. In Fig.3b, there are
satellite reflections along $b^{\ast}$ direction, in contrast to the
ED pattern for crystal 1, which shows no satellite reflections as
shown in Fig.3a . Because the XRD patterns of the two single
crystals have shown that crystal 1 is pure $n$=4 phase, while
crystal 2 is intergrowth of $n$=4 and $n$=3 phases, therefore, the
satellite reflections should come from the modulation structure in
$n$=3 phase. Superposition of the main reflections is consistent
with almost the same in-plane lattice parameters between two
compounds. Actually, the misfit ratio (1.62) in
[Ca$_2$CoO$_3$]$^{RS}$[CoO$_2$]$_{1.62}$ is slightly smaller than
that in crystal 1. But if Ca is partly substituted by Bi in
[Ca$_2$CoO$_3$]$^{RS}$[CoO$_2$]$_{1.62}$, the lattice parameter of
the RS layer is enhanced and the same misfit ratio could be obtained
as that in crystal 1 \cite{Mikami}.

The intergrowth of the $n$=3 and $n$=4 phases in Bi-Ca-Co-O system
arises from the thermodynamical competition for the two phases. Pure
single crystal of the $n$=4 phase can be grown with cooling the
melting solution from 900$\celsius$, while single crystal
intergrowth of $n$=3 and $n$=4 phases can be obtained with cooling
the same melting solution from 950$\celsius$. It definitely
indicates that the single crystal of $n$=3 phase can be grown only
above 900$\celsius$. The melting solution was slowly cooled from
950$\celsius$, the single crystal of $n$=3 phase starts to grow.
When the melting solution was cooled to below 900$\celsius$, the
$n$=4 phase is thermodynamically more stable than the $n$=3 phase.
Therefore, the single crystal of $n$=4 phase begins to grow with the
$n$=3 phase. In addition, the two phases have the same lattice
parameters in plane, which provides a condition for the epitaxial
intergrowth of the two phases. This is possible intergrowth
mechanism for the $n$=3 and $n$=4 phases in Bi-Ca-Co-O system.

\subsection{Physical Properties}

Temperature dependence of the in-plane resistivity for crystal 1 and
2 is plotted in Fig. 4. The room-temperature values of $\rho_{ab}$
are 12.7 m$\Omega$ cm for crystal 1 and 13.6 m$\Omega$ cm for
crystal 2, respectively. They are larger than those in
[Ca$_2$CoO$_3$]$^{RS}$[CoO$_2$]$_{1.62}$ (8.5 m$\Omega$) and
[Bi$_{0.87}$SrO$_2$]$_2^{RS}$[CoO$_2$]$_{1.82}$ (4.0 m$\Omega$)
\cite{Luo,Yamamoto}. $\rho_{ab}$ shows metallic behavior
(d$\rho$/d$T >$ 0) at high temperature and exhibits a minimum at 156
K and 206 K for crystal 1 and crystal 2, respectively. These
temperatures corresponding to the minimum of $\rho_{ab}$ ($T_{min}$)
are much higher than those observed in
[Ca$_2$CoO$_3$]$^{RS}$[CoO$_2$]$_{1.62}$ (70 K) and
[Bi$_{0.87}$SrO$_2$]$_2^{RS}$[CoO$_2$]$_{1.82}$ (80 K) single
crystals \cite{Luo,Yamamoto}. The room-temperature resistivity
$\rho_{ab}$ and the temperature corresponding to the minimum
resistivity $T_{min}$ indicate that Bi-Ca-Co-O system has the weaker
metallicity compared to [Ca$_2$CoO$_3$]$^{RS}$[CoO$_2$]$_{1.62}$ and
[Bi$_{0.87}$SrO$_2$]$_2^{RS}$[CoO$_2$]$_{1.82}$. Fig. 5 shows that
the insulator-like behavior at low temperature for the two types of
crystal follows different transport laws. Resistivity for crystal 1
below 150 K can be fitted using two thermal activation behavior
($\rho\propto exp[\Delta_g/k_BT]$, where $\Delta_g$ is the energy
gap), which gives the thermal activation energy ($\Delta_g$) to be
2.4 meV below 20 K and 4.3 meV between 25 K and 90 K. Resistivity
for crystal 2 can be fitted using thermal activation law above 40 K,
with $\Delta_g$=4.8 meV. Below 40 K, variable range hopping (VRH,
$\rho\propto exp[(T_{0}/T)^{1/4}]$) resistivity was observed.

Temperature dependence of the in-plane TEP is shown in the Fig.6 for
crystal 1 and crystal 2, as well as for
[Ca$_2$CoO$_3$]$^{RS}$[CoO$_2$]$_{1.62}$ crystal. \cite{Luo} The
magnitudes of 180 $\mu$V/K for crystal 1 and 190 $\mu$V/K for
crystal 2 are much larger than that observed in
[Ca$_2$CoO$_3$]$^{RS}$[CoO$_2$]$_{1.62}$ (125 $\mu$V/K) and Pb-doped
[Bi$_{0.87}$SrO$_2$]$_2^{RS}$[CoO$_2$]$_{1.82}$ (highest of 150
$\mu$V/K as x=0) \cite{Luo,Funahashi1}. This is consistent with the
weaker metallicity in Bi-Ca-Co-O system inferred by Fig. 4. The
in-plane TEP changes slightly at high temperature and decreases
obviously below 100 K, similar to that observed in
[Ca$_2$CoO$_3$]$^{RS}$[CoO$_2$]$_{1.62}$ and
[Bi$_{0.87}$SrO$_2$]$_2^{RS}$[CoO$_2$]$_{1.82}$. A model for the TEP
in layered cobaltites has be proposed by Koshibae \cite{Koshibae}.
The result for cobalt ions in low spin state is that the TEP depends
on the fraction of holes, \textit{x}=Co$^{4+}$/Co, according to the
expression derived from the generalized Heikes formula
\begin{equation}S=-\frac{k_B}{|e|}{\rm ln}(\frac{1}{6}\frac{x}{1-x})\end{equation}
where $k_B$ is the Boltzmann constant and $e$ is the charge of
electron. From this formula, the average valence of cobalt ions in
[Ca$_2$Bi$_{1.4}$Co$_{0.6}$O$_4$]$^{RS}$[CoO$_2$]$_{1.69}$ could be
3.42. The thermoelectric power factor ($Q=S^2/\rho$) calculated from
the data in Fig.6 and Fig.4 is shown in Fig.7. The two crystals have
almost the same value of $Q$=2.7$\times10^{-4}$ W/mK$^2$ at room
temperature, which is close to that obtained in Na$_x$CoO$_2$
\cite{Ohtaki}. However, Fig.7 shows that the room-temperature values
of $Q$ for crystal 1 with $n$=4 and crystal 2 with intergrowth is
obviously higher than that in
[Ca$_2$CoO$_3$]$^{RS}$[CoO$_2$]$_{1.62}$ single crystal with $n$=3.
Unchange in $Q$ at room temperature for crystal 2 with intergrowth
of $n$=3 and $n$=4 in Bi-Ca-Co-O system can be ascribed to the
simultaneous enhancement of TEP and resistivity. The enhancement of
the TEP together with the resistivity due to intergrowth in
Bi-Ca-Co-Co system is different from the case of the Sr-based system
\cite{Maignan4}, in which the intergrowth of $n$=3 and $n$=4 phase
enhances the TEP but reduces the resistivity relative to the pure
$n$=3 phase. Nevertheless, relative to pure $n$=3 phase, the power
factor is enhanced by the intergrowth of $n$=4 and $n$=3 phase in
Ca- and Sr-based system. It seems that the $n$=4 component plays the
major role in the transport properties in the single crystal with
intergrowth of $n$=4 and $n$=3 phase.

\section{CONCLUSION}
Single crystals of Bi-Ca-Co-O have been grown using K$_2$CO$_3$+KCl
flux. Single crystals with pure $n$=4 phase were grown from
900$\celsius$, while intergrowth of $n$=4 and $n$=3 phase was
obtained as crystals were grown from 950$\celsius$. At room
temperature, Ca-based $n$=4 crystal have the much larger TEP (180
$\mu$V/K) than that in Sr-based one and $n$=3
[Ca$_2$CoO$_3$]$^{RS}$[CoO$_2$]$_{1.62}$. The intergrowth of $n=4$
and $n=3$ phase enhances the TEP value and resistivity, but does not
change the power factor at room temperature. As pointed by Klein
\textit{et al.} \cite{Maignan4}, the presence of such intergrowth
could be an important structural feature of the misfit cobaltites in
generating larger values of TEP.

\section{ACKNOWLEDGEMENT}
This work is supported by the National Natural Science Foundation
of China and by the Ministry of Science and Technology of China
(973 project No: 2006CB601001).

\clearpage

\begin{figure}[htp]
\includegraphics[width=0.8\textwidth]{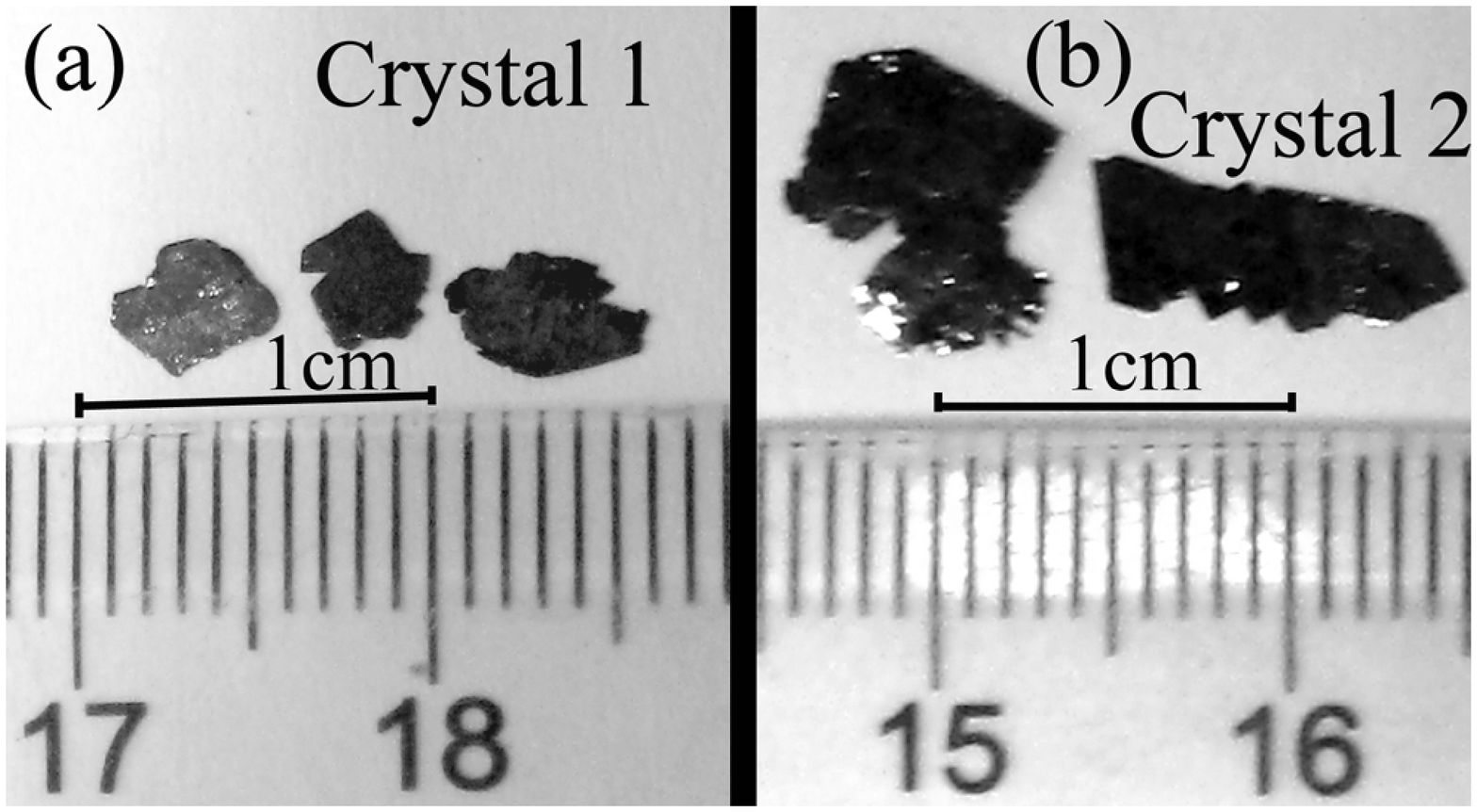}
\caption{Photographs of crystal 1 and 2.}
\end{figure}

\clearpage

\begin{figure}[htp]
\includegraphics[width=0.8\textwidth]{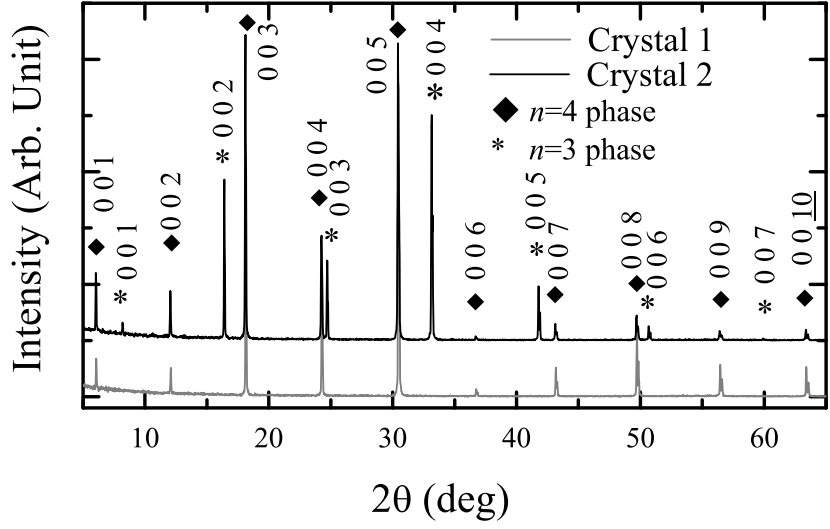}
\caption{The XRD patterns for crystal 1 (grown from 900 $\celsius$)
and crystal 2 (grown from 950$\celsius$). $\blacklozenge$ and $\ast$
are referred to the reflections from $n$=4 phase and $n$=3 phase,
respectively. Intergrowth can be clearly observed in the pattern for
crystal 2.}
\end{figure}

\clearpage

\begin{figure}[htbp]
\includegraphics[width=0.8\textwidth]{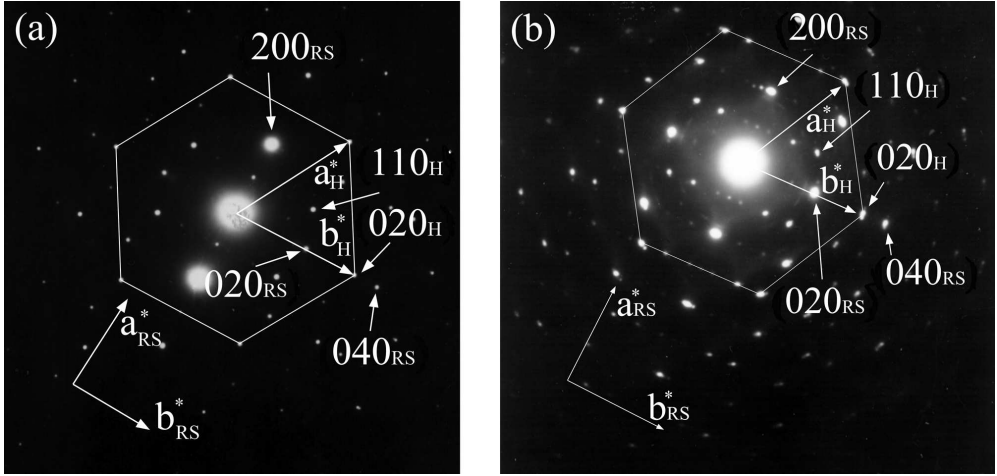}
\caption{The [001] ED patterns for the crystal 1 (a) and crystal 2
(b).}
\end{figure}

\clearpage

\begin{figure}[htp]
\includegraphics[width=0.8\textwidth]{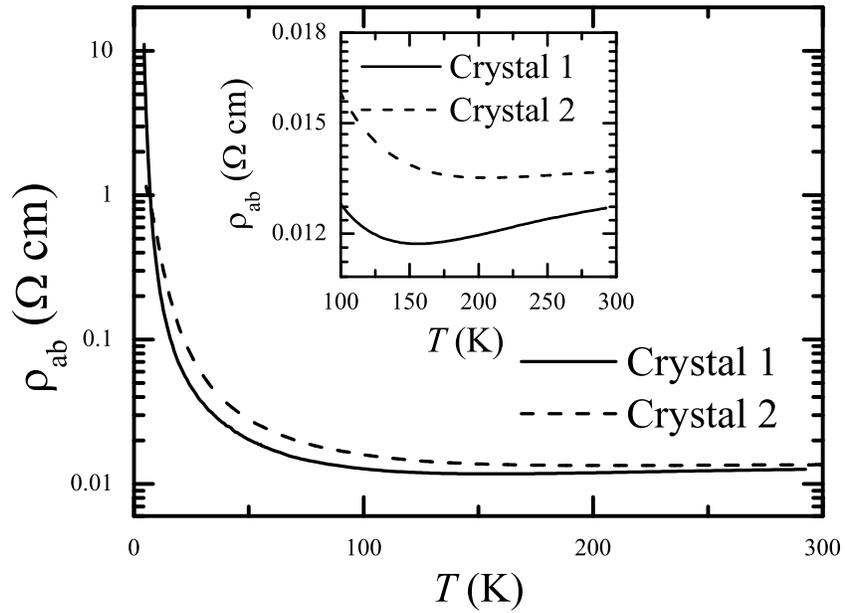}
\caption{Temperature dependence of in-plane resistivity for crystal
1 and 2. The inset shows the closeup of the in-plane resistivity at
high temperature, in which metallic behavior and minimum of
resistivity can be observed.}
\end{figure}

\clearpage

\begin{figure}[htp]
\includegraphics[width=0.8\textwidth]{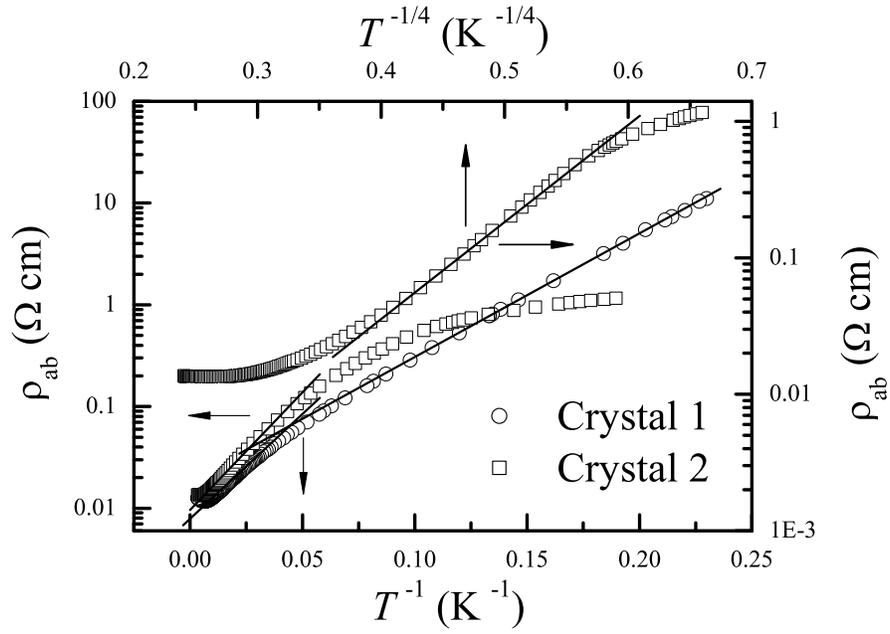}
\caption{Thermal activation ($\rho\propto exp[\Delta_g/k_BT]$, where
$\Delta_g$ is the energy gap) and variable range hopping
($\rho\propto exp[(T_{0}/T)^{1/4}]$) fitting of the resistivity for
crystal 1 and crystal 2.}
\end{figure}

\clearpage

\begin{figure}[htp]
\includegraphics[width=0.8\textwidth]{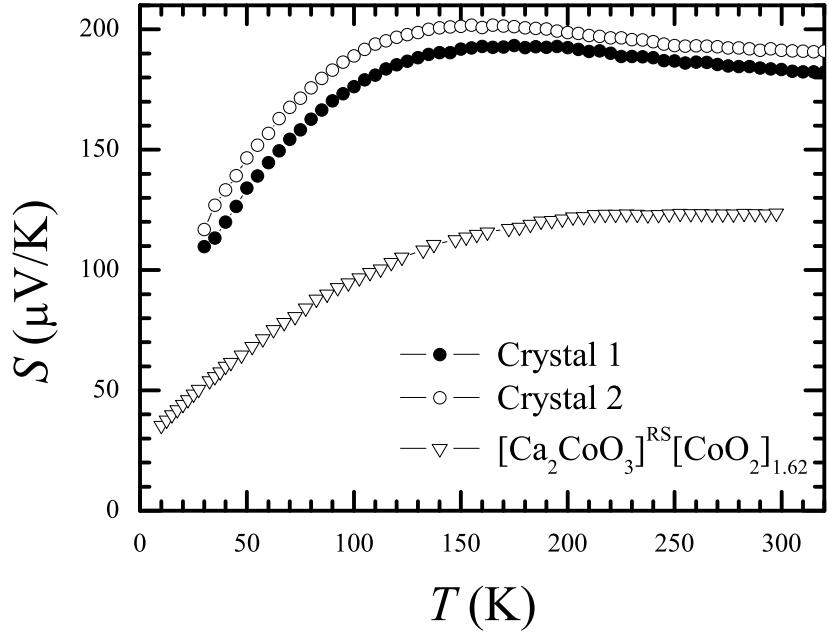}
\caption{Temperature dependence of Seebeck coefficient for crystal
1, crystal 2. That of [Ca$_2$CoO$_3$]$^{RS}$[CoO$_2$]$_{1.62}$
single crystal is also shown in the figure as comparison
\cite{Luo}.}
\end{figure}

\clearpage

\begin{figure}[htp]
\includegraphics[width=0.8\textwidth]{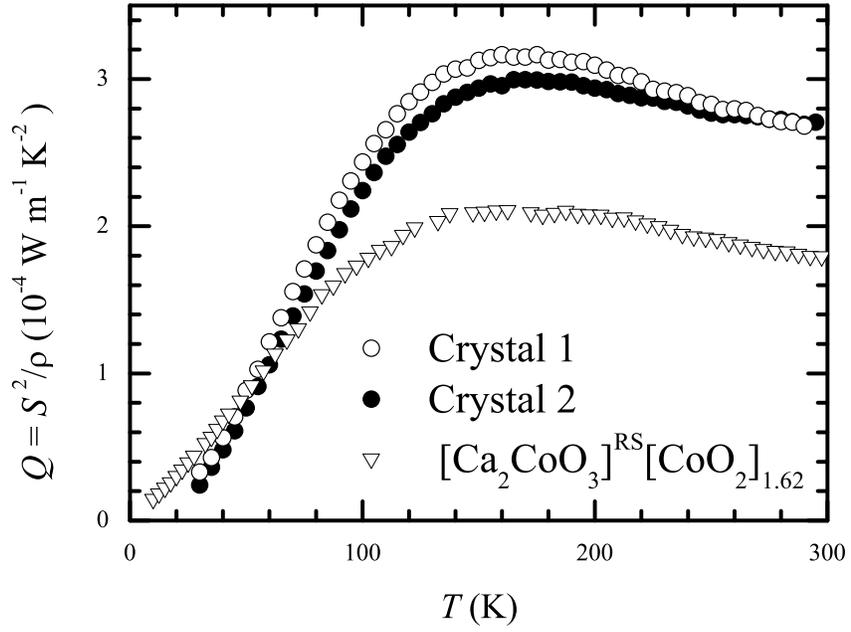}
\caption{Temperature dependence of power factor ($Q$) for crystal 1,
crystal 2, and [Ca$_2$CoO$_3$]$^{RS}$[CoO$_2$]$_{1.62}$ single
crystal \cite{Luo}.}
\end{figure}

\end{document}